# Calibrating evanescent-wave penetration depths for biological TIRF microscopy

*Short title:* TIRF image quantification

Martin Oheim, [*,†,‡,1 ✉], Adi Salomon, [¶,2] Adam Weissman, [¶] Maia Brunstein, [*,†,‡,§] and Ute Becherer[£]

[*] SPPIN – Saints Pères Paris Institute for the Neurosciences, F-75006 Paris, France;
[†] CNRS, UMR 8118, Brain Physiology Laboratory, 45 rue des Saints Pères, Paris, F-75006 France;
[‡] Fédération de Recherche en Neurosciences FR3636, Faculté de Sciences Fondamentales et Biomédicales, Université Paris Descartes, PRES Sorbonne Paris Cité, F-75006 Paris, France;
[¶]Department of Chemistry, Institute of Nanotechnology and Advanced Materials (BINA), Bar-Ilan University, Ramat-Gan, 5290002, Israel;
[§]*Chaire d'Excellence Junior*, Université Sorbonne Paris Cité, Paris, F-75006 France;
[£]Saarland University, Department of Physiology, CIPMM, Building 48, D-66421 Homburg/Saar, Germany;

[✉] Address all correspondence to
Dr Martin Oheim
SPPIN – Saints Pères Paris Institute for the Neurosciences
45 rue des Saints Pères
F-75006 Paris

Phone:　　+33 1 4286 4221 (Lab), -4222 (Office)
Fax:　　　+33 1 4286 3830
E-mails:　martin.oheim@parisdescartes.fr

[1)] MO is a Joseph Meyerhof invited professor with the Department of Biomolecular Sciences, The Weizmann Institute for Science, Rehovot, Israel.

[2)] AS was an invited professor with the Faculty of Fundamental and Biomedical Sciences, Paris Descartes University, Paris, France during the academic year 2017-18.

ABSTRACT. Roughly half of a cell's proteins are located at or near the plasma membrane. In this restricted space, the cell senses its environment, signals to its neighbors and exchanges cargo through exo- and endocytotic mechanisms. Ligands bind to receptors, ions flow across channel pores, and transmitters and metabolites are transported against concentration gradients. Receptors, ion channels, pumps and transporters are the molecular substrates of these biological processes and they constitute important targets for drug discovery. Total internal reflection fluorescence (TIRF) microscopy suppresses background from cell deeper layers and provides contrast for selectively imaging dynamic processes near the basal membrane of live-cells. The optical sectioning of TIRF is based on the excitation confinement of the evanescent wave generated at the glass/cell interface. How deep the excitation light actually penetrates the sample is difficult to know, making the quantitative interpretation of TIRF data problematic. Nevertheless, many applications like super-resolution microscopy, co-localization, FRET, near-membrane fluorescence recovery after photobleaching, uncaging or photo-activation/switching, as well as single-particle tracking require the quantitative interpretation of EW-excited images. Here, we review existing techniques for characterizing evanescent fields and we provide a roadmap for comparing TIRF data across images, experiments, and laboratories.

(193 words)

## Quantifying total internal reflection fluorescence

The purpose of TIRF is to selectively illuminate fluorophores that are right near a surface and not illuminate fluorophores that are further into the volume, above the surface. Major uses of TIRF include single-molecule detection (SMD) – e.g., for the localization-based super-resolution microscopies – as well as studies of cell-substrate contact region of living cells grown in culture. Typical biological applications are the investigation of cell adhesion sites, of the dynamics of membrane receptors, single-vesicle exo- and endocytosis or ER/plasma-membrane contact sites. The confinement of excitation light to a thin, near-interface layer results in background reduction and contrast enhancement.

Many TIRF applications require knowing the depth of the illuminated layer, e.g., for quantifying the motion of molecules or cellular organelles near the substrate. In principle, that depth can be calculated from the local refractive index (RI) of the medium ($n_1$) and the incidence angle of the illumination ($\theta$) at a known wavelength ($\lambda$). But in practice, the actual depth is much less certain. The incident light is often not perfectly collimated but in a range of angles, which sets up a continuous range of penetration depths. Also, the optics used to guide the incident beam upon the sample itself scatters light, some of which enters the sample as propagating light, which does not decay exponentially with distance from the TIR surface. Another source of uncertainty in TIR depth is that the sample's refractive index is not known precisely, and it may contain non-homogeneities, producing scattering.

In this perspective article, we briefly recall the fundamentals of evanescent waves (EWs) prior to discussing the concepts, methods and difficulties of calibrating TIRF intensities. We suggest a protocol for TIRF image quantification to better control, compare and share results. Other aspects of TIRF and TIRF microscopy have been covered in a number of excellent reviews and tutorials [1; 2; 3; 4; 5; 6; 7; 8].

## How to 'see' EWs?

Light impinging at an interface between two media[a] having, respectively, RIs $n_1$ and $n_3$ with $n_3 > n_1$ is partially reflected and partially transmitted. Snell's and Fresnel's laws govern, respectively, the angles and intensities of the reflected and refracted beams. A discontinuity is observed at a very oblique angle, above the critical angle $\theta_c = \mathrm{asin}(n_1/n_3)$ at which the reflected intensity becomes equivalent to that of the incident beam. Yet, there is light in the rarer medium ($n_1$), because energy and momentum conservation at the boundary prescribe the existence of a near field, skimming the surface. This evanescent (vanishing) wave propagates on the surface (as does the refracted beam for $\theta = \theta_c$) but its intensity is decaying exponentially perpendicular to the surface. Typical decay lengths are a fraction of the wavelength and can be as small as the $\lambda/5$, depending on the angle of incidence, $\theta$, **Fig. 1**$A$.

---

[a] an intermediate layer having a RI $n_2$ and thickness $d_2$ is for the moment neglected.

EWs are non-visible in the far field, but scattering or absorption couple out energy from the near field, measurable either as an intensity loss in the reflected beam ('attenuated' or 'frustrated' TIR), or as fluorescence excitation in a thin boundary layer above the reflecting interface (TIRF). EW scattering at RI-discontinuities at or near the interface results in far-field light, too. In either case, the total intensity depends on how deep the EW reaches into the rare medium ($n_1$).

Another way to 'see' EWs is by their coupling to collective electron oscillations ('plasmons') in a thin metal film ($n_2$, $d_2$) deposited on the glass. Momentum matching is only possible due to the foreshortening of the EW wave vector (wave-front squeezing) and it leads to a sharp decrease in the reflected intensity at the surface plasmon resonance (SPR) angle, $\theta_{SPR} > \theta_c$. Changes in $n_3$, e.g., by the binding of molecules to the surface, will shift $\theta_{SPR}$. This angle shift is the basis for SPR-based sensing, spectroscopy, and SPR microscopy of cell/substrate contact sites [9; 10; 11].

(**Figure 1** *Light confinement by an evanescent wave* close to here)

**The theoretical framework**

For a beam impinging from the left, the EW propagates along $+x$. Its intensity decays in axial ($z$-) direction, with – in theory – an exponential dependence, $I(z;\theta) = I_0(\theta) \exp[-z/\delta(\theta)]$. The distance over which the intensity drops to $1/e$ (37%) of its value at the interface ($z = 0$) is called the 'penetration depth',

$$\delta(\theta) = \frac{\lambda}{4\pi\sqrt{n_3^2\sin^2(\theta)-n_1^2}} = \frac{\lambda}{4\pi n_3\sqrt{\sin^2(\theta)-\sin^2(\theta_c)}} \qquad (\text{Eq. 1})$$

It depends on the excitation wavelength $\lambda$, the polar beam angle $\theta$ and the RIs, **Fig. 1**$A$. We here omitted possible intermediate layers ($n_j$, $d_j$) [12; 13]. The penetration depth, $\delta$, does neither depend on the polarization nor on the azimuthal angle $\phi$ of the incoming beam. On a plot of $\delta(\theta)$ vs. $\theta$, **Fig. 1**$B$, we recognize an asymptotic behavior, both for $\theta \to \theta_c$, for which $\delta$ diverges and becomes infinite (which is intuitive, because of the emergence of the transmitted, refracted beam for $\theta < \theta_c$), and for very large $\theta \to \pi/2$, for which $\delta(\theta)$ approaches $\lambda/(4\pi n_3 \sqrt{1 - \sin^2(\theta_c)})$, **Fig. 1**C. For a given $\lambda$, higher-index substrates ($n_3$) or more grazing angles $\theta$ result in a better optical sectioning.

**Many unknowns affect the true penetration depth**

As there is a smooth intensity roll-off, the actual penetration depth is strictly defined only in relation to a certain signal-to-background or signal-to-noise level. For a mono-exponential, $I(z)$ decays to <5% of $I(0)$ over an axial distance of two penetration depths, $z = 2\delta$. With some arbitrariness, one could thus take $2\delta$ as the effective probe depth, because 5% is of the order of typical noise levels. Unfortunately, the question of how far excitation light reaches into the cell is more complicated, and it has been a matter of passionate debate (see, e.g., http://lists.umn.edu/cgi-bin/wa?A2=ind1011&L=confocalmicroscopy&D=0&P=18718). In practice, $\delta$ is only known within certain bounds, and often even the assumption of a single-exponential intensity decay does not hold. The reasons are the following:

(*i*), whereas $\lambda$ and $n_3$ are identified, neither $n_1$ nor $\theta$ are known with much precision. For a biological cell, $n_1$ varies on a microscopic length-scale [14] modifying both $\theta_c$ and $\delta$, **Fig. 1**C. But also $\theta$ is only known with a certain accuracy (adjustment accuracy) and precision (beam divergence). These unknowns translate into an uncertainty and range of penetration depths $\delta(\theta)$, respectively.

(*ii*), even with all parameters known, the calculated penetration depth is just that: theoretical, because microscope- and sample-generated non-evanescent light modifies the intensity decay that will no longer be a simple exponential. For prismless TIRF [15] several reports have shown that excitation light distribution is best described by the superposition of a rapidly decaying and a long-range component that is fairly independent of $\theta$ [16; 17; 18]. The effect of this long-range component is paradoxical: while most of the excited fluorescence is due to EW-excitation for $\theta$ just above $\theta_c$, for very high incidence angles (that normally should produce better optical sectioning) non-evanescent light dominates.

(*iii*), background comes from stray excitation from inside the microscope objective and different optical surfaces [17]. In the critical illumination scheme used in multi-angle TIRF, any scatter on the scanning mirrors (or in any conjugate sample plane) is imaged into the sample plane, **Figs. 2**A and **2**B, (*a*). Although this stray excitation can be quantitated [16; 17] such measurements are not routine;

(*iv*), protein-rich adhesion sites or near-membrane organelles like lysosomes or dense-core granules have a higher RI than the surrounding cytoplasm. The spatial inhomogeneity in $n_1(x, y, z)$ does not only affect $\delta$ directly (eq.1), but it also produces non-evanescent light, **Fig. 2***B,* panel (*b*). Scattering occurs predominantly in forward direction, into angles close to the original propagation direction of the EW [17; 19; 20]. For even higher local RI $\geq n_1$, light can be refracted, generating intense beams in EW propagation direction, as observed in chromaffin cells packed with secretory granules [19], **Fig. 2***B*, (*c*).

In view of these difficulties, authors and microscope manufacturers have resorted to calculating $\delta(\theta)$ using eq.(1) – see https://imagej.nih.gov/ij/plugins/tirf/index.html for a popular ImageJ plug-in – leading to overly optimistic if not unrealistic statements of light confinement at the reflecting interface. The interpretation of TIRF intensities in terms of fluorophore concentration changes, fluorophore axial distances or single-particle trajectories is thus often flawed by large error bars or, worse, simply wrong.

## Azimuthal beam spinning does not improve axial confinement

At least for illumination non-homogeneities there is remedy. A straightforward solution is 2-photon EW excitation [20; 21; 22]. Another is azimuthal beam scanning [23; 24; 25]: as scattering is directional, varying the EW propagation direction scrambles both propagation and scattering directions and reduces the flare in any given direction, **Fig. 2***C*. 'Spinning' TIRF (spTIRF) or *incoherent* ring illumination [26] reduces interference fringes and illumination non-uniformities but it does not change the problem of the presence of non-evanescent excitation light [17] as it only *redistributes* and equalizes intensities, **Fig. 2***C*. spTIRF is increasingly being used [17; 27; 28; 29; 30; 31; 32; 33; 34; 35] and commercial systems are available (e.g., from TILL: 'Polytrope' [36], Leica [37], Roperscientific iLas [34]), however, the bulk of published TIRF images has been acquired with unidirectional illumination.

(**figure 2** *Illumination imperfections and azimuthal beam-scanning TIRF* close to here)

## Why calibrating penetration depths at all?

There are a number of biophysical techniques that do *not* just aim at background rejection but make quantitative use of TIRF intensities, either for *localizing* fluorophores or for measuring axial *concentration profiles*:

- *VA-TIRF*. Variable-angle TIRF (VA-TIRF) is a technique that uses systematic variations of $\theta$ for reconstructing axial fluorophore distributions, **Fig. 3**A. For a given axial fluorophore profile $C(z)$ the fluorescence $F(\theta)$ displays a characteristic shape. With $\theta$ known and $F(\theta)$ measured, $C(z)$ can be obtained, pixel by pixel by an inversion procedure [19; 38; 39; 40; 41; 42; 43; 44]. VA-TIRF relies on the precise knowledge of the shape of the axial intensity decay. Many studies assumed a mono-exponential intensity decay [45; 46; 47; 48; 49], which perhaps holds for prism-type TIRF but seems overly optimistic for objective-type TIRF [16; 17; 33].

- *TIRF-colocalization*. Knowledge of the axial intensity decay is mandatory for multi-color excitation, **Fig. 3**B. As the penetration depth scales linearly with $\lambda$ (eq.1), one can compensate the $\lambda$-dependence of $\delta$ by adjusting $\theta$ to maintain a constant probe volume in different color channels [34; 35; 50; 51]. The same applies to controls in TIR-FRET [31] with direct and FRET excitation of donor fluorescence.

- *TIRF-FRAP* [52], **Fig. 3**C, uses the combination of localized EW-photobleaching and TIRF imaging of the fluorescence recovery after photobleaching to study near-interface fluorophore mobility. The interpretation of FRAP data relies on the known probe volume. $\delta(\theta)$ calibration is even more stringent for bleaching- or photoswitching-based axial super-resolution measurements based upon consecutive VA-TIRF imaging of deeper and fluorescence deletion in more proximal layers [53]. Analogous arguments hold for TIRF photoactivation and photoswitching experiments, which include the growing group of PALM/STORM super-resolution studies, as well as optogenetic activation using EW-illumination [54].

- *TIR-FCS*. TIRF correlation spectroscopy [55; 56; 57] follows the same principles as confocal-spot FCS but gains sensitivity and surface selectivity from the additional excitation confinement, particularly when combined with confocal-spot TIRF excitation [58; 59].

- Even for less specialized TIRF applications a known penetration depth is a pre-requisite for reproducibility and comparing data among experiments, laboratories and publications.

(**figure 3** *Quantitative uses of TIRF* close to here)

In the sequel, we review techniques for measuring $\theta$ and then calibrating $\delta(\theta)$ and we discuss their respective benefits and problems. We also comment on supercritical angle fluorescence (SAF) microscopy [60; 61], or "evanescence in emission" [4; 62; 63] and how it can be used either as an axial nanoscale ruler [64; 65] or be combined with TIRF to achieve a better near-membrane confinement than obtained with TIRF alone [33]. Finally, we describe a simple yet effective way for calibrating the true optical sectioning of the microscope, based on the acquisition of combined TIRF and EPI *z*-image stacks of a thin 3-D sample of sub-diffraction beads embedded in a low-melting point agarose gel that mimics the refractive index of the cytoplasm [66].

**Measuring the polar beam angle**

Most techniques for calibrating $\delta$ rely on determining $\theta$ and measuring the intensity resulting from a known axial fluorophore distribution $F(z)$. A minimal $\theta$ calibration can be obtained by measuring the positions of the setscrew or stepper motor at three characteristic points: (*i*), normal incidence, $\theta = 0$, epifluorescence (EPI), (*ii*), the critical angle $\theta_c$ and, (*iii*), the limiting angle $\theta_{NA}$ of the objective. Identified either from sample-plane [20] or back-focal plane (BFP) images [17; 25; 48; 67], Abbe's sine condition, $r = f_{obj} n_1 \sin\theta$, is used to extrapolate to intermediate values. $r$ is the radius in the exit pupil plane (objective BFP), and $f$ is the focal length of the objective, $f_{obj} = f_{TL}/M$, with $f_{TL}$ being the focal length of the manufacturer tube lens, and $M$ is the objective transversal magnification, **Fig. 4***A*. For a VA- or spTIRF setup, one can substitute $r$ using the focal length of the focusing lens, $f_{FL}$, and the substrate refractive index $n_1$ to obtain

$$\theta = \sin^{-1}\left[\frac{f_{TL}}{f_{obj}} \cdot \frac{\sin(k \cdot U_\theta)}{n_1}\right], \quad \text{(Eq.2)}$$

and $k$ is a constant (°/V) characteristic for the scanner used, and $U_\theta$ is the voltage applied to the polar axis of the scanner.

Alternatively, one can measure $\theta$ in the sample plane, with the advantage that coverslip tilt and beam misalignment are accounted for. In the 'lateral-displacement' technique, the laser is set to an oblique angle and the coverslip is topped with a drop of dye solution or a thin flurophore film. Defocus produces a lateral movement of the fluorescent spot that can be traced from fitting a 2-Gaussian with the elliptical cross-section of the beam, **Fig. 4***B*. Fitting a straight line with its center position yields arcsin(*θ*) [18; 68]. Alternatively, for the beam not to suffer TIR, the glass/air surface can be made transmissive by an oil-coupled external prism [25; 35] or a solid-immersion lens (SIL) [17], **Fig. 4***C*, and the (refracted) beam is projected to the ceiling or wall. In a variant of this triangulation technique, the back-reflected beam is picked up and projected onto a quadrant photodiode to determine *θ* from the center-of-mass of a Gaussian fitted with the beam profile [69]. Similarly, taking out the emission filter permits to see the back-reflected light on a BFP image, and calculating $\theta = \arcsin[r/(n_1 f_{\mathrm{obj}})]$ [67], **Fig. 4***D*.

(**figure 4** *Polar beam-angle calibration* close to here)

**Intensity-based measurements of evanescent-wave penetration depths**

Once a look-up table for *θ* has been generated, we can adjust and estimate *δ*(*θ*). This involves localizing the reflecting interface and measuring the fluorescence (or some other variable) for different fluorophore heights. This procedure is repeated for several incidence angles *θ* and the obtained curve compared with the calculated one (fig. 1B).

*a) Calibration samples with a known axial fluorophore profile.* **Fig. 5** shows examples of such samples, including the "raisin cake", a random, sparse 3-D distribution of sub-diffraction fluorescent microspheres in an index-matched gel, **Fig. 5***A*. The acquisition of a *z*-stack of images can localize these beads with respect to the reflecting interface [66]. Alternatively, point emitters can be fixed to the surface of an oblique microscope slide, **Fig. 5***B*, [34], a large convex [70] or concave lens [67]. Equivalently, one can use the contour of an index-matched dye-coated large-diameter fluorescent bead, or an unlabeled bead embedded in dye-containing medium [16], **Fig. 5***C*. An elegant variant is the use of a tilted fluorescently labeled microtubule [71].

All the above test samples have in common the requirement for an evenly lit field of view (see [17] for the limits of this approximation) and they all need *z*-scans to locate the fluorophores with respect to the reflecting interface. This adds a complication for objective-type TIRF as the reflected beam displaces laterally upon focusing (see above), and the *z*-dependence of the PSF (detection volume) and *z*-dependent spherical aberrations require for a correction, see [71] and our calibration protocol, below.

*b) Semi-infinite dye layers.* Simpler is the use of VA-TIRF and a thick, $d \gg \delta(\theta)$, homogenous fluorescent sample [20; 40] to estimate the effective penetration depth from the variation of the *cumulative* fluorescence, **Fig. 5***D*. Assuming a two-component axial decay, i.e., the sum of mono-exponentially decaying EW with a decay length $\delta(\theta)$ and a long-range component with $D \gg \delta$ [16; 18] we express the measured fluorescence as

$$F(z) = b + A \cdot e^{-z/\delta(\theta)} + B \cdot e^{-z/D} \qquad \text{(Eq.3)}$$

Here, we assumed that $D$ to be only slowly varying with $\theta$ and the offset $b$ negligible. After integration over $z$ in the bounds $[0, \infty]$, eq.3 yields a linear dependence of the measured fluorescence on $\delta(\theta)$,

$$F_{tot} = A \cdot \delta(\theta) \cdot \left(1 - e^{-\frac{z}{\delta(\theta)}}\right) + B \cdot D \cdot \left(1 - e^{-\frac{z}{D}}\right) \approx A \cdot \delta(\theta) + B, \qquad \text{(Eq.4)}$$

because the second term is an angle-independent offset. If eq.4 is normalized for the $\theta$-dependence of the incident intensity at the interface, $I_0(z=0)$, e.g., by recording $F_0(\theta)$ of a thin fluorophore film at the interface [20; 42; 72], then the implicit $\theta$-dependence of $A$ and $B$ is cancelled out. For the integral to solve as shown one has to assume that the $\theta$-dependent term of the intensity decay follows a mono-exponential.

*c) Fluorescence correlation spectroscopy (FCS).* Another way to determine the axial confinement is TIR-FCS [55; 56; 57]. The intensity fluctuations resulting from single molecules moving in and out of the excitation volume are being used to estimate its size, **Fig. 5***E*. Assuming a mono-exponential intensity decay, one can explicitly solve the autocorrelation function and back-calculate the dye concentration, the bulk diffusion coefficient and the excitation volume [55]. Complications arise from dye adsorption and hindered mobility at the interface, from the distance-dependent fluorescence collection efficiency [40] as a consequence of the change in the fluorophore radiation pattern for interface-proximal dipoles (see below) and of a non-exponential axial intensity decay.

*d) Single-spot measurements.* Other approaches measure $\delta(\theta)$ only in one point, either by sampling the *local* EW intensity with a thinned optical fiber tip connected to a photodetector [73], or by measuring the fluorescence generated by a sub-resolution fluorescent bead fixed at the tip of an AFM cantilever [74; 75; 76] or the tip of a micropipette [45]. Brutzer *et al*.

used a four-arm DNA junction as a nanomechanical translation stage to propel a single fluorescent quantum dot through EW field [77], **Fig. 5**F. A similar strategy uses a combination of magnetic tweezers and a supercoiling DNA together with a fluorescent nanodiamond-labeled magnetic bead and surface-immobilized fluorescent nanodiamonds as a reference for $z = 0$ [78]. Alternatively, fluorophores at different distances $z$ could be obtained with static 3-D scaffolds, e.g., tetrahedral DNA-origami fluorescent rulers of different dimensions [79; 80; 81].

(**figure 5** close to here)

*e) Index-matched polymer steps.* Many of the above approaches either perturb the EW by the presence of RI boundaries (i.e., the edge of a glass slide, a lens touching the interface, or a fiber tip), they require tedious sample preparation, or they are not applicable in the aqueous environment of biological TIRF. Taniguchi's group introduced a calibration slide featuring steps of different nanometric heights fabricated from a non-fluorescent polymer, the RI of which matched that of water. This spacer staircase was topped with sub-diffraction fluorescent microspheres generating fluorescent steps at different distances from the reflecting interface [82; 83], **Fig. 6**A, *left*. The Schwille lab recently proposed a similar strategy using - instead of combining several nanofabrication techniques as in the Unno papers - a simpler dip-coating method. They topped their staircase polymer with AlexaFLuor488 solution and measured the EW-excited fluorescence as a function of step height [18], reminiscent of the "infinite dye layer" technique, **Fig. 6**A, *right*. Corroborating earlier work, both studies confirmed that the axial intensity profile contained both evanescent and non-evanescent contribution and that the effective penetration depth exceeded the calculated one.

A yet different approach uses thin films of optically transparent polymers on glass substrates wrapping a nanometric dye layer. **Fig. 6**B illustrates such a 'sandwich' consisting of a non-fluorescent spacer layer that onto which J-aggregates were electrostatically adsorbed, resulting in a ~2-nm thin, homogenous emitter layer, **Fig. 6**B, *inset*. This layer was finally spin cast with another, 5-µm thick, polymer layer [84]. The advantage of this multi-layer test sample is that it produces a thin, controlled and homogenous dye distribution at a precisely controlled distance from the reflecting interface. With several of such test samples, plotting the collected fluorescence as a function of the fluorophore distance Δ allows the measurement of the axial effective intensity decay, **Fig. 6**C.

(**figure 6** close to here)

**Other descriptors than intensity**

Until now, we have focused on techniques for calibrating the EW that relied on fluorescence intensity measurements from test samples having a known axial fluororphore profile. However, other parameters of fluorescence can be used, too.

*Fluorophore radiation pattern.* Fluorophores change their radiation pattern when they approach a RI boundary closer than ~$\lambda$ because the evanescent component of their radiation can couple to the interface, become propagative [85; 86; 87] and detectable in the far field at angles >$\theta_c$ 'forbidden' by Snell's law. The selective detection of this 'super-critical angle fluorescence' (SAF) [62; 88; 89] features a similar surface selectivity and background suppression of TIRF whilst not requiring EW excitation. As SAF probes the same sub-$\lambda$ length-scale, it can be used in conjunction with EW-excitation for achieving a better surface selectivity than TIRF alone [33].

Fourier-plane imaging conveniently measures the radiation pattern [90]. The BFP image has other benefits for calibrated TIRF microscopy: it allows determining the cell's near-membrane RI, $n_1$, from the radius at which the transition between SAF and undercritical angle fluorescence (UAF) occurs [14; 63; 91]. The SAF/UAF intensity ratio is proportional to fluorophore height [4; 63; 64; 91] a feature, which can be used for combined axial fluorophore localization and penetration-depth calibration by plotting the total fluorescence SAF+UAF *vs*. (SAF/UAF) [84] (Fig. 6C).

*Fluorescence lifetime*, $\tau$ is a measure of how long the molecule stays in the excited state after absorption of a photon. $\tau$ is one over the sum of the radiative and non-radiative decay rates. Fluorescence lifetime oscillations and shortening are observed in the presence of a metal coating [92; 93], a near-field probe tip [94] or a metal nanoparticle [95]. For the sub-$\lambda$ distances relevant here, non-radiative energy transfer from the excited molecule to the metal offers an alternative decay path for excited-state relaxation. Distance-dependent surface quenching by a thin gold layer was used to measure fluorophore heights of dye-labeled microtubules [96] on the basis of the model by Chance, Prock, and Silbey (CPS model) that relates $\tau$ to the fluorophore height $z$ [92].

While the distance-dependent lifetime quenching is strong for metal, only a small effect is observed on bare glass [86]. Also, other factors than surface distance interfere with molecular lifetimes, including the orientation [97], RI [98], solvent polarity, viscosity and by complex formation and collisions in the presence of nearby quenchers.

*Other*, more exotic ways for axially localizing particles and calibrating $\delta(\theta)$ include interferometric, PSF-engineering-based, axially-structured illumination-based or multi-plane based axial super-resolution techniques (reviewed in [99]), but these generally require considerably more complex instrumentation. Yet other approaches use not light but other physical parameters that are modified by the presence of an interface, e.g., the anisotropic diffusion near an interface [100] or the flow velocity gradient of particles moving under the influence of a rotating disk [101] to calibrate the EW decay.

In summary, experimenters dispose of a host of observables for estimating the axial intensity decay at or near a dielectric interface. In spite of a common quest for quantitative TIRF imaging, neither a consensual test sample, nor common metrics, nor an industry standard have emerged for calibrating evanescent-wave excited fluorescence intensities. Instead, different labs have come up with their own customized solutions, making the comparison among studies difficult, if not impossible. In the now dominant prismless objective-type TIRF, the quantitative interpretation of TIRF intensities is problematic due to the co-existence of a localized, evanescent and a non-evanescent, long-range excitation component that has consistently been observed using different protocols [16; 17; 18]. Furthermore, at the same beam angle and beam diameter, the same TIRF objective, used in conjunction with different illumination optics, will perform differently. It would thus seem important that the field agrees on a protocol for testing, evaluating and quantifying axial confinement in EW techniques.

**A simple protocol for interpreting, comparing, and sharing TIRF data**

A reference standard for calibrating the effective probe depth in TIRF and SAF microscopies should meet the following requirements, it *(i)*　　　should work in an aqueous environment mimicking the cytoplasm; specifically, it should not perturb the EW by introducing objects of a different refractive index,

*(ii)*　　　should not require modifications to existing TIRF and SAF setups,

*(iii)*　　should be easily transposable from one lab to the other, be stable or, alternatively, easily and reproducibly to fabricate,

*(iv)*　　should mimic the conditions at the dielectric interface typically encountered in biological TIRF microscopy, i.e., it should reproduce the refractive index of the intracellular environment, and,

*(v)*　　　should be compatible with different TIRF geometries to permit comparisons among setups and experiments.

We propose the 'raisin cake' method for measuring the effective penetration depth in a sparse 3-D fluorophore distribution. The method is based on commercially available fluorescent microspheres located on the glass/water interface and at different axial positions in a transparent, index-matched agarose gel [102]. It does not require any specific equipment other than that available in any wet lab. However, in as much as the calibration procedure involves the acquisition of *z*-stacks of images, a precise focus control for accurate and nanometric focusing is needed.

100-nm TetraSpecks$^{TM}$ beads were deposited on a coverslip and at different axial distances from the coverslip by embedding them in a low-density agarose solution containing sucrose to increase the refractive index to 1.374 (see Appendix for details). We first took an epifluorescence (EPI) image stack across the sample in which the individual image planes were spaced by Δ*z* = 10 nm, covering a range from -1.0 μm below to +2 μm above the reflecting interface. Next we acquired for the focal position corresponding to the plane in which the TIRF image was sharp ($z_0$), one EPI and three TIRF images, varying the polar beam angle *θ* between each TIRF acquisition. We used spTIRF to even out illumination heterogeneities, **Fig. 7***A*. Acquisitions were realized with autofocus feedback to avoid focal drift during recordings. We determined the axial location of each bead by measuring the background-subtracted intensity in a 3×3 pixel region of interest (ROI) centered on each bead and plotted it against *z*. **Fig. 7***B* shows examples of two beads located at different axial positions. We fitted Gaussian distributions with the axial fluorophore profiles and chose the peak location of the Gaussian as the bead position. Similar to the sub-diffraction *xy*-localization precision in PALM or dSTORM, our method allowed us to determine bead positions with 15-nm precision, well below the depth-of-field of the objective. 100-nm diameter beads were found to give more consistent results that 200-nm beads. However, even for the smaller beads the axial intensity profile did not always follow a Gaussian distribution. In some cases impurities or microbubbles in the agarose, low signal-to-noise or the superimposition of 2 beads distorted the profile, **Fig. 7**B, *right*. Other errors arise from beads that are directly attached to the coverslip because the vicinity to the glass/medium interface distorts the axial intensity profile due to spherical aberration at the RI-boundary, **Fig. 7**B, *right*. We therefore eliminated all beads that displayed a non-Gaussian profile from analysis. Finally, the peak of the Gaussian distribution on the EPI *z*-stack was taken as the intrinsic fluorescence of the beads, **Fig. 7**C), and used to normalize their fluorescence.

(**figure 7** *Raisin-cake calibration* close to here)

The EPI image acquired at $z_0$ allowed us to assess the influence of the depth-of-focus of the objective on the fluorescence intensity of beads remote from the interface. Keeping the focus at $z_0$, the EPI intensity of the off-focus beads decayed with distance to the focal plane with a length constant of 481 ± 166 nm, **Fig.7**D. The same intensity envelope will also modulate the bead intensities in TIRF and be convoluted with the EW excitation intensity decay. Indeed, axial intensity decay of the beads upon TIRF was best described by the superposition of a long-range and short-range exponential decay (Eq.3). We set the long-range component to the earlier measured EPI decay ($D$ = 481 ± 166 nm) whereas the short-range component was modulated by changing $\theta$. Focusing at the TIRF-illuminated layer ($z_0$) we extracted the fluorescence intensity of all beads in the field of view at different polar angles, ranging from just above $\theta_c$ to $\theta_{NA}$. For each bead, the obtained fluorescence intensity was normalized to the intrinsic (in-focus) EPI fluorescence of the bead and plotted against its *z*-position and a double exponential fitted with the ensemble of data points. We took the short-range component as the effective penetration depth $\delta(\theta)$. At $\theta$ = 67.5°, 70.0 and 72.5°, $\delta(\theta)$ was (139 ± 20) nm, (109 ± 16) nm, and (91 ± 13) nm, respectively, close to the calculated penetration depths of 145, 111 and 94 nm, respectively, given by the software by the iLAS module (Gataca systems, France). At $\theta$ = 65°, close to $\theta_c$ = asin(1.375/1.52) =64.8°, fitting a single- or double exponential with the measured fluorescence intensity decay similarly resulted in $\delta(65°)$ = 367 ± 217 or 349 ± 118 nm, respectively, indicating that non-evanescent excitation light indistinguishable from EPI dominated and impaired the penetration depth estimation, compared to a calculated $\delta_{calc}(65°) = 308$ nm. Thus, similar to earlier reports [16; 25; 33 Niederauer, 2018 #67] [18], we find that supercritical-angle illumination through the periphery of a high-NA objective produces a mix of evanescent and EPI illumination. For beam angles of 72.5°, 70.0 and 67.5°, the fractional weight $B/(b+A+B)$ from eq.(3) of non-evanescent excitation light amounted to 13, 15.3, and 15.5%. However, to which degree of evanescent vs. non-evanescent light a fluorophore is exposed to, will depend on its very distance from the reflecting interface and, paradoxically, the contribution of non-evanescent light much exceeds the above percentages for $z \gg 0$.

Our technique permits a reliable measurement of the effective axial confinement at the reflecting interface. However, all depends on the axial-localization accuracy of the beads, which in turn is a function of signal. The same protocol is applicable to fixed cells, in which small organelles are labeled (not shown). In this case, the density of labeled organelles has to be low to avoid the superimposition of fluorescent structures. Furthermore, the object size and their fluorescent intensity have to be fairly homogenous to avoid artifacts (see Appendix).

**Conclusions**

Prism-based TIRF can provide 'cleaner' and more uniform illumination as well as a higher-accuracy angular positioning, therefore it should be applied where the acquisition of an accurate intensity profile is of importance. Point-scanning TIRF (albeit slower) has slightly better lateral resolution than wide-field TIRF and can be combined with SAF detection to improve axial confinement. Independent of the very geometry used, variety of EW-calibration technique exists, among which we recommend the simple, inexpensive and reproducible 'raisin-cake' technique.

- first, calibrate the polar beam angle $\theta$, either by ray tracing through the illumination optical path or via triangulation after the passage of the beam through the objective;
- Use SAF refractometry with the same fluorophores you plan to use for TIRF imaging to get an idea of the average refractive index, $<n_1>$, surrounding the label;
- Use polar beam angles not to close to the critical angle;
- Use $\theta$-sweeps to study how different angles affect the TIRF image;
- Use azimuthal beam spinning TIRF to homogeneously illuminate the scene;

- Calibrate the effective penetration depth using the 'raisin cake' technique;
- Systematically report $\theta \pm \Delta\theta$, $<n_1 \pm \Delta n_1>$, and the estimated $\delta \pm \Delta\delta$;
- Combine TIRF excitation and *v*SAF detection for better optical sectioning;
- Discuss, how the uncertainty in $\delta$ and TIRF intensities affects your conclusions.

## APPENDIX A: TIRF test-sample protocol

*Sample preparation.* Small droplets of a 1:4 dilution in water of 0.1-µm diameter Tetraspeck™ beads (Invitrogen) were deposited on a glass coverslip (BK-7, e.g., FisherScientific or Marienfeld) and allowed to dry, **Fig. S1***A*. The bead surplus was flushed with water. In parallel, a solution of 0.5% low-melting point agarose solution (seaplaque GTG agarose, Lonza) containing 842 mM sucrose was prepared. Sucrose increased the RI of the solution to 1.375 to mimic the refractive index of cells [103]. The solution was heated to ~80°C in a water bath to melt the agarose. It was then kept at constant ~50°C. 5 µl of sonicated beads were transferred to a pre-heated Eppendorf tube to which 15 µl agarose/sucrose solution was added and thoroughly triturated with a preheated pipette. The now 20 µl of solution was pipetted onto the coverslip at the same location, topping the beads that were previously attached to the glass (**Fig. S1***B*). The agarose was allowed to polymerize. To speed up this step the coverslip can be placed on an ice-cold surface. Then, 1-2 ml agarose/sucrose solution (not containing beads) was slowly deposited on top of the bead-containing drop to achieve a thick layer of RI = 1.375. To obtain good results, it is important that this solution is maintained at a temperature just above the gelling point of agarose (~30°C) so that the small bead-containing drop already on the coverslip does not re-melt when the large volume of agarose/sucrose solution is added.

*Setup.* The penetration depth was assessed on a commercial multi-angle objective-type TIRF setup (Visitron Systems, Puchheim, Germany) built around an IX83 inverted microscope equipped with an autofocus module, a UAPON100XOTIRF NA-1.49 objective (all Olympus). Precise focusing was achieved with the motorized focus ($\Delta z$ = 10 nm). The beam of a 488-nm 100-mW laser was directed into the iLAS$^2$ beam scanning system (Gataca Systems, France) allowing ($\theta,\phi$) scans. Fluorescence was collected through the same objective, extracted with a ZT405/488/561/640rpc multi-band dichroic, a ZET405/488/561/640rpc multi-band emission filter (both from Semrock) and detected on an Evolve-EM515 EMCCD camera (Photometrics). All setup components were controlled by VisiView (Visitron Systems GmbH). The resulting pixel size in the sample plane was 160 nm. Typical integration times were 100 ms at gain 1 and an EM gain of 200 abitrary units.

*Data analysis*

Images were analyzed with imageJ (Rasband, W.S., ImageJ, U. S. National Institutes of Health, Bethesda, Maryland, USA, http://imagej.nih.gov/ij/, 1997-2014). Fitting and further analysis of the data was performed with IGOR PRO software (Wavemetrics, Lake Oswego, OR, USA).

## Acknowledgements

Recent work in our labs related to TIRF and SAF microscopy was financed by the *Agence Nationale de la Recherche* (ANR-10-INSB-04-01, *grands investissements* FranceBioImaging, FBI, to MO), and a Chaire d'Excellence Junior Université Sorbonne Paris Cité (USPC) to MB, and the Université Paris Descartes (invited professorship during the academic year 2017-18, to AS). MO and AS acknowledge support from a French-Israeli CNRS-WIS ImagiNano LIA grant. UB acknowledges funding from the *Deutsche Forschungsgemeinschaft* (DFG, CRC894). The Oheim lab is a member of the C'nano IdF and *Ecole de Neurosciences de Paris* (ENP) excellence clusters for nanobiotechnology and neurosciences, respectively.

## Author contributions

MO, UB, MB, AS, and AW performed experiments, MB, UB and AW prepared samples, MO, UB and AS analyzed the data, MO wrote the manuscript with contributions from all authors.
All authors have given their approval to the final version of the manuscript.

## Conflicting interest statement

The authors declare no conflict of interest.

## List of abbreviations

| | | |
|---|---|---|
| DNA | - | desoxy-ribonucleic acid |
| EPI | - | epifluorescence |
| EGFP | - | enhanced green fluorescent protein |
| EW | - | evanescent wave |
| FCS | - | fluorescence correlation spectroscopy |
| FRAP | - | fluorescence recovery after photobleaching |
| NA | - | numerical aperture |
| PALM | - | photoactivation localization microscopy |
| ROI | - | region of interest |
| SAF | - | supercritical angle fluoresence |
| SIL | - | solid immersion lens |
| SIM | - | structured illumination microscopy |
| SMD | - | single molecule detection |
| SOFI | - | superresolution optical fluctuation imaging |
| SPR | - | surface plasmon resonance |
| spTIRF | - | spinning TIRF |
| STED | - | stimulated emission depletion |
| STORM | - | stochastic optical reconstruction microscopy |
| TIR(F(M)) | - | total internal reflection (fluorescence (microscopy)) |
| VA-TIRF | - | variable-angle TIRF |

# **REFERENCES**

[1] H. Schneckenburger, Total internal reflection fluorescence microscopy: technical innovations and novel applications. Curr Op Biotechnol 16 (2005) 13-18.

[2] K.N. Fish, Total internal reflection fluorescence (TIRF) microscopy. Curr. Prot. Cytometry (2009) 12.18. 1-12.18. 13.

[3] A.L. Mattheyses, S.M. Simon, and J.Z. Rappoport, Imaging with total internal reflection fluorescence microscopy for the cell biologist. J. Cell Sci. 123 (2010) 3621-3628.

[4] D. Axelrod, Evanescent excitation and emission in fluorescence microscopy. Biophys. J. 104 (2013) 1401-9.

[5] M.L. Martin-Fernandez, C.J. Tynan, and S.E. Webb, A 'pocket guide' to total internal reflection fluorescence. J. Microsc. 252 (2013) 16-22.

[6] N.S. Poulter, W.T. Pitkeathly, P.J. Smith, and J.Z. Rappoport, The physical basis of total internal reflection fluorescence (TIRF) microscopy and its cellular applications. Methods Mol Biol 1251 (2015) 1-23.

[7] M. Oheim, TIRF (Total Internal Reflection Fluorescence). eLS (2016) DOI: 10.1002/9780470015902.a0022505.

[8] L.J. Young, F. Ströhl, and C.F. Kaminski, A guide to structured illumination TIRF microscopy at high speed with multiple colors. JoVE (2016).

[9] K.-F. Giebel, C. Bechinger, S. Herminghaus, M. Riedel, P. Leiderer, U. Weiland, and M. Bastmeyer, Imaging of cell/substrate contacts of living cells with surface plasmon resonance microscopy. Biophys. J. 76 (1999) 509-516.

[10] V. Yashunsky, V. Lirtsman, M. Golosovsky, D. Davidov, and B. Aroeti, Real-time monitoring of epithelial cell-cell and cell-substrate interactions by infrared surface plasmon spectroscopy. Biophys. J. 99 (2010) 4028-4036.

[11] T. Son, J. Seo, I.-H. Choi, and D. Kim, Label-free quantification of cell-to-substrate separation by surface plasmon resonance microscopy. Optics Commun. 422 (2018) 64-68.

[12] W. Reichert, and G. Truskey, Total internal reflection fluorescence (TIRF) microscopy. I. Modelling cell contact region fluorescence. J Cell Sci 96 (1990) 219-230.

[13] D. Axelrod, Fluorescence excitation and imaging of single molecules near dielectric-coated and bare surfaces: a theoretical study. Journal of microscopy 247 (2012) 147-160.

[14] M. Brunstein, L. Roy, and M. Oheim, Near-membrane refractometry using supercritical angle fluorescence. Biophys. J. 112 (2017) 1940-1948.

[15] A.L. Stout, and D. Axelrod, Evanescent field excitation of fluorescence by epi-illumination microscopy. Appl. Opt. 28 (1989) 5237-5242.

[16] A.L. Mattheyses, and D. Axelrod, Direct measurement of the evanescent field profile produced by objective-based total internal reflection fluorescence. J. Biomed. Opt. 11 (2006) 014006-014006-7.

[17] M. Brunstein, M. Teremetz, K. Hérault, C. Tourain, and M. Oheim, Eliminating unwanted far-field excitation in objective-type TIRF. Part I. Identifying sources of nonevanescent excitation light. Biophys. J. 106 (2014) 1020-1032.

[18] C. Niederauer, P. Blumhardt, J. Mücksch, M. Heymann, A. Lambacher, and P. Schwille, Direct characterization of the evanescent field in objective-type total internal reflection fluorescence microscopy. Opt. Express 26 (2018) 20492-20506.

[19] A. Rohrbach, Observing secretory granules with a multiangle evanescent wave microscope. Biophys. J. 78 (2000) 2641-2654.

[20] F. Schapper, J.T. Gonçalves, and M. Oheim, Fluorescence imaging with two-photon evanescent wave excitation. Eur. Biophys. J. 32 (2003) 635-643.

[21] M. Oheim, and F. Schapper, Non-linear evanescent-field imaging. J. Phys. D: Appl. Phys. 38 (2005) R185.
[22] R.S. Lane, A.N. Macpherson, and S.W. Magennis, Signal enhancement in multiphoton TIRF microscopy by shaping of broadband femtosecond pulses. Opt. Express 20 (2012) 25948-25959.
[23] A.L. Mattheyses, K. Shaw, and D. Axelrod, Effective elimination of laser interference fringing in fluorescence microscopy by spinning azimuthal incidence angle. Microsc. Res. Tech. 69 (2006) 642-647.
[24] R. Fiolka, Y. Belyaev, H. Ewers, and A. Stemmer, Even illumination in total internal reflection fluorescence microscopy using laser light. Microsc. Res. Tech. 71 (2008) 45-50.
[25] M. van't Hoff, V. de Sars, and M. Oheim, A programmable light engine for quantitative single molecule TIRF and HILO imaging. Opt. Express 16 (2008) 18495-18504.
[26] D. Axelrod, Cell-substrate contacts illuminated by total internal reflection fluorescence. J Cell Biol 89 (1981) 141-145.
[27] D.S. Johnson, J.K. Jaiswal, and S. Simon, Total internal reflection fluorescence (TIRF) microscopy illuminator for improved imaging of cell surface events. Curr. Prot. Cytometry (2012) 12.29. 1-12.29. 19.
[28] S. Abdelhady, S.S. Kitambi, V. Lundin, R. Aufschnaiter, P. Sekyrova, I. Sinha, K.T. Lundgren, G. Castelo-Branco, S. Linnarsson, R. Wedlich-Soldner, A. Teixeira, and M. Andang, Erg channel is critical in controlling cell volume during cell cycle in embryonic stem cells. PLoS One 8 (2013) e72409.
[29] A.H. Crevenna, N. Naredi-Rainer, A. Schonichen, J. Dzubiella, D.L. Barber, D.C. Lamb, and R. Wedlich-Soldner, Electrostatics control actin filament nucleation and elongation kinetics. J. Biol. Chem. 288 (2013) 12102-13.
[30] P.-I. Ku, Anna K. Miller, J. Ballew, V. Sandrin, Frederick R. Adler, and S. Saffarian, Identification of Pauses during Formation of HIV-1 Virus Like Particles. Biophys. J. 105 (2013) 2262-2272.
[31] J. Lin, and A.D. Hoppe, Uniform total internal reflection fluorescence illumination enables live cell fluorescence resonance energy transfer microscopy. Microsc. Microanal. 19 (2013) 350-359.
[32] P.-I. Ku, M. Bendjennat, J. Ballew, M.B. Landesman, and S. Saffarian, ALIX Is Recruited Temporarily into HIV-1 Budding Sites at the End of Gag Assembly. PLoS ONE 9 (2014) e96950.
[33] M. Brunstein, K. Hérault, and M. Oheim, Eliminating unwanted far-field excitation in objective-type TIRF. Part II. Combined evanescent-wave excitation and supercritical-angle fluorescence detection improves optical sectioning. Biophys. J. 106 (2014) 1044-1056.
[34] J. Boulanger, C. Gueudry, D. Münch, B. Cinquin, P. Paul-Gilloteaux, S. Bardin, C. Guérin, F. Senger, L. Blanchoin, and J. Salamero, Fast high-resolution 3D total internal reflection fluorescence microscopy by incidence angle scanning and azimuthal averaging. Proc. Acad. Sci. USA 111 (2014) 17164-17169.
[35] W. Zong, X. Huang, C. Zhang, T. Yuan, L.-l. Zhu, M. Fan, and L. Chen, Shadowless-illuminated variable-angle TIRF (siva-TIRF) microscopy for the observation of spatial-temporal dynamics in live cells. Biomed. Opt. Express 5 (2014) 1530-1540.
[36] J. Riedl, A.H. Crevenna, K. Kessenbrock, J.H. Yu, D. Neukirchen, M. Bista, F. Bradke, D. Jenne, T.A. Holak, Z. Werb, M. Sixt, and R. Wedlich-Söldner, Lifeact: a versatile marker to visualize F-actin. Nat. Methods 5 (2008) 605-7.
[37] T. Veitinger, Controlling the TIRF Penetration Depth is Mandatory for Reproducible Results. http://www.leica-microsystems.com/science-lab/controlling-the-tirf-penetration-depth-is-mandatory-for-reproducible-results/ Online (2012).

[38] W. Reichert, P. Suci, J. Ives, and J. Andrade, Evanescent detection of adsorbed protein concentration-distance profiles: fit of simple models to variable-angle total internal reflection fluorescence data. Appl. Spectrosc. 41 (1987) 503-508.

[39] J.S. Burmeister, G.A. Truskey, and W.M. Reichert, Quantitative analysis of variable-angle total internal reflection fluorescence microscopy (VA-TIRFM) of cell/substrate contacts. J Microsc 173 (1994) 39-51.

[40] B.P. Ölveczky, N. Periasamy, and A. Verkman, Mapping fluorophore distributions in three dimensions by quantitative multiple angle-total internal reflection fluorescence microscopy. Biophys J 73 (1997) 2836-2847.

[41] M. Oheim, D. Loerke, B. Preitz, and W. Stuhmer, Simple optical configuration for depth-resolved imaging using variable-angle evanescent-wave microscopy, Optical biopsies and microscopic techniques III, International Society for Optics and Photonics, 1999, pp. 131-141.

[42] D. Loerke, B. Preitz, W. Stuhmer, and M. Oheim, Super-resolution measurements with evanescent-wave fluorescence-excitation using variable beam incidence. Journal of biomedical optics 5 (2000) 23-31.

[43] K. Stock, R. Sailer, W.S. Strauss, M. Lyttek, R. Steiner, and H. Schneckenburger, Variable-angle total internal reflection fluorescence microscopy (VA-TIRFM): realization and application of a compact illumination device. J Microsc 211 (2003) 19-29.

[44] J. Li, W. Han, Y. Li, Y. Chen, Y. Shang, Y. Chen, and Z. Gui, Inverse problem based on the fast alternating direction method of multipliers algorithm in multiangle total internal reflection fluorescence microscopy. Appl. Opt. 57 (2018) 9828-9834.

[45] M. Oheim, D. Loerke, W. Stühmer, and R.H. Chow, The last few milliseconds in the life of a secretory granule. Eur. Biophys. J. 27 (1998) 83-98.

[46] C.D. Byrne, A.J. de Mello, and W.L. Barnes, Variable-angle time-resolved evanescent wave-induced fluorescence spectroscopy (VATR-EWIFS): a technique for concentration profiling fluorophores at dielectric interfaces. The Journal of Physical Chemistry B 102 (1998) 10326-10333.

[47] M. van't Hoff, M. Reuter, D.T. Dryden, and M. Oheim, Screening by imaging: scaling up single-DNA-molecule analysis with a novel parabolic VA-TIRF reflector and noise-reduction techniques. PhysChemChemPhys 11 (2009) 7713-7720.

[48] M.C. Dos Santos, R. Déturche, C. Vézy, and R. Jaffiol, Axial nanoscale localization by normalized total internal reflection fluorescence microscopy. Opt. Lett. 39 (2014) 869-872.

[49] M.C. Dos Santos, R. Déturche, C. Vézy, and R. Jaffiol, Topography of cells revealed by variable-angle total internal reflection fluorescence microscopy. Biophys J 111 (2016) 1316-1327.

[50] M. Oheim, and W. Stuhmer, Multiparameter evanescent-wave imaging in biological fluorescence microscopy. IEEE J Quant Elec 38 (2002) 142-148.

[51] U. Becherer, T. Moser, W. Stühmer, and M. Oheim, Calcium regulates exocytosis at the level of single vesicles. Nat Neurosci 6 (2003) 846.

[52] E. Pryazhnikov, D. Fayuk, M. Niittykoski, R. Giniatullin, and L. Khiroug, Unusually strong temperature dependence of P2X3 receptor traffic to the plasma membrane. Fronti. Cell. Neurosci. 5 (2011) 27.

[53] Y. Fu, P.W. Winter, R. Rojas, V. Wang, M. McAuliffe, and G.H. Patterson, Axial superresolution via multiangle TIRF microscopy with sequential imaging and photobleaching. Proc. Acad. Sci. USA 113 (2016) 4368-4373.

[54] D. Li, K. Hérault, E.Y. Isacoff, M. Oheim, and N. Ropert, Optogenetic activation of LiGluR-expressing astrocytes evokes anion channel-mediated glutamate release. J. Physiol. 590 (2012) 855-873.

[55] S. Harlepp, J. Robert, N. Darnton, and D. Chatenay, Subnanometric measurements of evanescent wave penetration depth using total internal reflection microscopy combined with fluorescent correlation spectroscopy. Applied physics letters 85 (2004) 3917-3919.

[56] K. Hassler, M. Leutenegger, P. Rigler, R. Rao, R. Rigler, M. Gösch, and T. Lasser, Total internal reflection fluorescence correlation spectroscopy (TIR-FCS) with low background and high count-rate per molecule. Optics Express 13 (2005) 7415-7423.

[57] N.L. Thompson, X. Wang, and P. Navaratnarajah, Total internal reflection with fluorescence correlation spectroscopy: Applications to substrate-supported planar membranes. J. Struct. Biol. 168 (2009) 95-106.

[58] T. Ruckstuhl, and S. Seeger, Attoliter detection volumes by confocal total-internal-reflection fluorescence microscopy. Opt. Lett. 29 (2004) 569-571.

[59] M. Leutenegger, C. Ringemann, T. Lasser, S.W. Hell, and C. Eggeling, Fluorescence correlation spectroscopy with a total internal reflection fluorescence STED microscope (TIRF-STED-FCS). Opt. Express 20 (2012) 5243-5263.

[60] T. Ruckstuhl, and D. Verdes, Supercritical angle fluorescence (SAF) microscopy. Opt. Express 12 (2004) 4246-4254.

[61] T. Barroca, K. Balaa, J. Delahaye, S. Lévêque-Fort, and E. Fort, Full-field supercritical angle fluorescence microscopy for live cell imaging. Opt. Lett. 36 (2011) 3051-3053.

[62] D. Axelrod, Selective imaging of surface fluorescence with very high aperture microscope objectives. J. Biomed. Opt. 6 (2001) 6-13.

[63] M. Brunstein, A. Salomon, and M. Oheim, Decoding the information contained in the fluorophore radiation pattern ACS Nano 12 (2018) 11725–11730.

[64] C.M. Winterflood, T. Ruckstuhl, D. Verdes, and S. Seeger, Nanometer axial resolution by three-dimensional supercritical angle fluorescence microscopy. Phys. Rev. Lett. 105 (2010) 108103.

[65] J. Deschamps, M. Mund, and J. Ries, 3D superresolution microscopy by supercritical angle detection. Opt. Express 22 (2014) 29081-29091.

[66] A. Quintana, C. Kummerow, C. Junker, U. Becherer, and M. Hoth, Morphological changes of T cells following formation of the immunological synapse modulate intracellular calcium signals. Cell Calcium 45 (2009) 109-122.

[67] E. Soubies, S. Schaub, A. Radwanska, E. Van Obberghen-Schilling, L. Blanc-Féraud, and G. Aubert, A framework for multi-angle TIRF microscope calibration, Biomedical Imaging (ISBI), 2016 IEEE 13th International Symposium on, IEEE, 2016, pp. 668-671.

[68] T.P. Burghardt, Measuring incidence angle for through-the-objective total internal reflection fluorescence microscopy. J. Biomed. Opt. 17 (2012) 126007-126007.

[69] D.S. Johnson, R. Toledo-Crow, A.L. Mattheyses, and S.M. Simon, Polarization-controlled TIRFM with focal drift and spatial field intensity correction. Biophys. J. 106 (2014) 1008-1019.

[70] J.A. Steyer, and W. Almers, Tracking single secretory granules in live chromaffin cells by evanescent-field fluorescence microscopy. Biophys J 76 (1999) 2262-2271.

[71] C. Gell, M. Berndt, J. Enderlein, and S. Diez, TIRF microscopy evanescent field calibration using tilted fluorescent microtubules. J. Microsc. 234 (2009) 38-46.

[72] M. Guo, P. Chandris, J.P. Giannini, A.J. Trexler, R. Fischer, J. Chen, H.D. Vishwasrao, I. Rey-Suarez, Y. Wu, and X. Wu, Single-shot super-resolution total internal reflection fluorescence microscopy. Nature methods 15 (2018) 425.

[73] A.J. Meixner, M.A. Bopp, and G. Tarrach, Direct measurement of standing evanescent waves with a photon-scanning tunneling microscope. Appl. Opt. 33 (1994) 7995-8000.

[74] A. Sarkar, R.B. Robertson, and J.M. Fernandez, Simultaneous atomic force microscope and fluorescence measurements of protein unfolding using a calibrated evanescent wave. Proc. Acad. Sci. USA 101 (2004) 12882-12886.

[75] J. Oreopoulos, and C.M. Yip, Combined scanning probe and total internal reflection fluorescence microscopy. Methods 46 (2008) 2-10.

[76] M. Franken, C. Poelma, and J. Westerweel, Nanoscale contact line visualization based on total internal reflection fluorescence microscopy. Optics express 21 (2013) 26093-26102.

[77] H. Brutzer, F.W. Schwarz, and R. Seidel, Scanning evanescent fields using a pointlike light source and a nanomechanical DNA gear. Nano letters 12 (2011) 473-478.

[78] Y. Seol, and K.C. Neuman, Combined Magnetic Tweezers and Micro-mirror Total Internal Reflection Fluorescence Microscope for Single-Molecule Manipulation and Visualization, Single Molecule Analysis, Springer, 2018, pp. 297-316.

[79] C. Steinhauer, R. Jungmann, T.L. Sobey, F.C. Simmel, and P. Tinnefeld, DNA origami as a nanoscopic ruler for super-resolution microscopy. Angew. Chem. 48 (2009) 8870-8873.

[80] J.J. Schmied, A. Gietl, P. Holzmeister, C. Forthmann, C. Steinhauer, T. Dammeyer, and P. Tinnefeld, Fluorescence and super-resolution standards based on DNA origami. Nat. Meth. 9 (2012) 1133.

[81] R. Schreiber, J. Do, E.-M. Roller, T. Zhang, V.J. Schüller, P.C. Nickels, J. Feldmann, and T. Liedl, Hierarchical assembly of metal nanoparticles, quantum dots and organic dyes using DNA origami scaffolds. Nat Nanotech 9 (2014) 74.

[82] N. Unno, A. Maeda, S.-i. Satake, T. Tsuji, and J. Taniguchi, Fabrication of nanostep for total internal reflection fluorescence microscopy to calibrate in water. Microelectronic Engineering 133 (2015) 98-103.

[83] N. Unno, H. Kigami, T. Fujinami, S. Nakata, S.-i. Satake, and J. Taniguchi, Fabrication of calibration plate for total internal reflection fluorescence microscopy using roll-type liquid transfer imprint lithography. Microelectronic Engineering 180 (2017) 86-92.

[84] M. Oheim, and A. Salomon, Calibration standard for evanescence microscopy. in: *e.a.* Centre National de la Recherche Scientifique - CNRS, (Ed.), 2018.

[85] C. Carniglia, L. Mandel, and K. Drexhage, Absorption and emission of evanescent photons. J. Opt. Soc. Am. 62 (1972) 479-486.

[86] W. Lukosz, and R. Kunz, Light emission by magnetic and electric dipoles close to a plane interface. I. Total radiated power. JOSA 67 (1977) 1607-1615.

[87] J. Mertz, Radiative absorption, fluorescence, and scattering of a classical dipole near a lossless interface: a unified description. JOSA B 17 (2000) 1906-1913.

[88] J. Enderlein, T. Ruckstuhl, and S. Seeger, Highly efficient optical detection of surface-generated fluorescence. Appl. Opt. 38 (1999) 724-732.

[89] T. Ruckstuhl, J. Enderlein, S. Jung, and S. Seeger, Forbidden light detection from single molecules. Anal. Chem. 72 (2000) 2117-2123.

[90] A.L. Mattheyses, and D. Axelrod, Fluorescence emission patterns near glass and metal-coated surfaces investigated with back focal plane imaging. J. Biomed. Opt. 10 (2005) 054007.

[91] N. Bourg, C. Mayet, G. Dupuis, T. Barroca, P. Bon, S. Lécart, E. Fort, and S. Lévêque-Fort, Direct optical nanoscopy with axially localized detection. Nat. Photonics 9 (2015) 587.

[92] R. Chance, A. Prock, and R. Silbey, Lifetime of an emitting molecule near a partially reflecting surface. J. Chem. Phys. 60 (1974) 2744-2748.

[93] K. Tews, On the variation of luminescence lifetimes. The approximations of the approximative methods. J. Luminesc. 9 (1974) 223-239.

[94] W.P. Ambrose, P.M. Goodwin, R.A. Keller, and J.C. Martin, Alterations of single molecule fluorescence lifetimes in near-field optical microscopy. Science 265 (1994) 364-367.

[95] J. Seelig, K. Leslie, A. Renn, S. Kühn, V. Jacobsen, M. van de Corput, C. Wyman, and V. Sandoghdar, Nanoparticle-induced fluorescence lifetime modification as nanoscopic ruler: demonstration at the single molecule level. Nano letters 7 (2007) 685-689.

[96] M. Berndt, M. Lorenz, J. Enderlein, and S. Diez, Axial Nanometer Distances Measured by Fluorescence Lifetime Imaging Microscopy. Nano Lett. 10 (2010) 1497-1500.

[97] M. Kreiter, M. Prummer, B. Hecht, and U. Wild, Orientation dependence of fluorescence lifetimes near an interface. J. Chem. Phys. 117 (2002) 9430-9433.
[98] S. Strickler, and R.A. Berg, Relationship between absorption intensity and fluorescence lifetime of molecules. J. Chem. Phys. 37 (1962) 814-822.
[99] W. Liu, K.C. Toussaint Jr, C. Okoro, D. Zhu, Y. Chen, C. Kuang, and X. Liu, Breaking the Axial Diffraction Limit: A Guide to Axial Super-Resolution Fluorescence Microscopy. Laser Photon. Rev. 12 (2018) 1700333.
[100] S. Saffarian, and T. Kirchhausen, Differential evanescence nanometry: live-cell fluorescence measurements with 10-nm axial resolution on the plasma membrane. Biophys. J. 94 (2008) 2333-2342.
[101] C. Zettner, and M. Yoda, Particle velocity field measurements in a near-wall flow using evanescent wave illumination. Exp. Fluids 34 (2003) 115-121.
[102] A. Quintana, C. Schwindling, A.S. Wenning, U. Becherer, J. Rettig, E.C. Schwarz, and M. Hoth, T cell activation requires mitochondrial translocation to the immunological synapse. Proc. Acad. Sci. USA 104 (2007) 14418-23.
[103] X. Liang, A. Liu, C. Lim, T. Ayi, and P. Yap, Determining refractive index of single living cell using an integrated microchip. Sensors Actuators A: Physical 133 (2007) 349-354.

## FIGURE LEGENDS

FIGURE 1. *Fundamentals of TIRF excitation*. (A), *left*, prismless (objective-type) TIRF. A laser beam is focused to an eccentric position in the back-focal plane (BFP, *dashed*) of a high-numerical aperture (NA) objective generating a collimated beam impinging at an oblique angle $\theta$ at the dielectric interface (*solid grey*, $n_3>n_1$). *Middle*, for $\theta$ exceeding the critical angle, $\theta_c = \operatorname{asin}(n_1/n_3)$, the beam is totally reflected at the interface and an inhomogeneous surface wave is generated in the rare medium ($n_1$). This 'evanescent' wave (EW) propagates along the surface (the Pointing vector, **S**, is oriented in +x direction for a beam impinging from the left, *red arrow*), and its intensity decays exponentially with axial (+z) distance from the reflecting interface, *right*, with a length constant ('penetration depth') $\delta$ of the order of 100 nm. (B), dependence of $\delta$ on $\theta$, for $\lambda = 488$ nm, $n_1 = 1.35$, and for different substrate indices $n_3$. The higher $n_3$ the smaller the critical angle $\theta_c$ and the better the optical sectioning. For a typical borosilicate glass/cell-interface and a NA-1.45 objective, the maximally attainable angle $\theta_{NA}$ limits the penetration depth to 73 nm (*red dash*). In this angle range, $\delta$ depends steeply on $\theta$, demanding high precision and accuracy when adjusting $\theta$. (C), dependence of $\theta_c$ and $\delta_\infty$ on substrate index, $n_3$. The asymptotes of the critical angle $\theta_c$ and limiting penetration depth $\delta_\infty$ for grazing incidence ($\theta \rightarrow 90°$), respectively, decrease monotonously with $n_3$. *Red* line indicates $n_3 = 1.52$ (BK-7), as before.

FIGURE 2. *Azimuthal beam scanning evens out illumination imperfections*. (A), layout for polar- and azimuthal beam scanning TIRF ('spinning' TIRF, spTIRF). A 2-axis scan ($\theta'$, $\phi'$) is relayed via a *4f* beam compressor, BC, into an equivalent sample plane, ESP, and - via the telescope formed by the focusing lens, FL, and objective (obj) – imaged into the sample plane (SP). *Inset* shows light distribution in the back-focal plane, BFP. Abbreviations: dic – dichroic mirror, EMCCD – electron-multiplying charged-coupled device camera, EBFP – equivalent back-focal plane, SD – scanning device. Due to this critical illumination any illumination imperfections in ESP' and ESP show up in SP. (B), sources of non-evanescent excitation, (*a*), dust on the scanning mirrors and in intermediate sample planes is directly imaged into the sample plane, resulting in illumination impurities and glare; (*b*) EW scattering at refractive-index (RI) boundaries produces light propagating in forward direction, modifying the effective $\delta(\theta)$ across the field-of-view; (*c*), for shallow angles $\theta \gtrsim \theta_c$, protein-rich cell adhesion sites can have a RI high enough to disrupt total reflection and generate intense beams of refracted light. (C), negative-staining experiment, in which a non-labeled BON cell was embedded in fluorescein-dextran containing extracellular saline and imaged in unidirectional TIRF (*top*) and spTIRF (*bottom*), respectively (from Brunstein *et al.* BJ 2014a). The bottom of the cell adhering to the coverslip excludes the extracellular dye and appears as a dark 'footprint'. Note the flare in EW propagation direction (*top*), which is abolished upon spTIRF (*bottom*).

FIGURE 3. *Quantitative uses of TIRF*. (A), variable-angle (VA-) TIRF. Smaller $\theta$ translate into larger illumination depths $\delta(\theta)$ (block arrow) allowing a topographic reconstruction of axial fluorophore profiles from a multi-$\theta$ stack. (B), multi-$\lambda$ excitation. $\delta$ scales linearly with $\lambda$ (block arrows). Toggling between different excitation wavelengths alters the excited volume (grey

arrowheads), *top*. Adjusting $\theta$ between different-color acquisitions maintains a constant excitation volume, permitting quantitative co-localization or FRET studies at or near the basal plasma membrane, *bottom*. (C), TIR-FRAP, in this variant of fluorescence recovery after photobleaching, a sample is sequentially imaged, bleached and re-imaged with EW excitation. An intense pulse of evanescent light (flash) selectively bleaches the surface-proximal fluorophores. Unbleached molecules from deeper sample regions repopulate the bleached volume. TIRF-FRAP allows studying the average mobility and mobile fraction of near-membrane fluorescent species, *inset*. Here, the penetration depth is constant, and the illumination intensity is modulated between imaging and bleaching episodes.

FIGURE 4. *Incidence-angle calibration*. (A), the radial position $r$ of a focused laser spot in the BFP unambiguously determines the beam angle $\theta$. Three characteristic points can be easily identified: (1), epifluorescence (EPI) at normal incidence, $\theta = 0 \Leftrightarrow r = 0$; (2) the disappearance of the refracted beam at $r_c$, $\Leftrightarrow \theta_c$ (TIR) and, (3) the disappearance of the reflected beam at $r_{NA}$, when the incident beam is beyond the limiting numerical aperture of the objective $\theta_{NA} = \arcsin(NA/n_1)$. Intermediate values of $\theta(r)$ are interpolated from Abbe's sine condition, $r = f_{obj}\, n_1 \sin(\theta)$. (B), lateral-displacement assay. The beam is directed at an oblique angle against an oil-coupled coverslip, coated with a thin fluorophore layer, *green*. Depending on the amount of defocus, the beam intersects the coverslip at different positions, and the lateral offset of the fluorescent spot on the camera image, together with knowledge of the piezo-controlled defocus permits to triangulate $\theta$. A similar strategy is used in (C) for $\theta > \theta_c$. Here, an oil-coupled, index-matched solid immersion lens (SIL) prohibits TIR and couples out a refracted beam that is then projected against the wall or some ruler at large distance to triangulate $\theta$. (D), the positional information contained in the reflected beam is used to measure $r(\theta)$ by coupling out a small percentage of the reflected intensity with a miniature semitransparent micromirror ($\mu$) in the periphery of the objective and direct it onto a position-sensitive detector (PSD), e.g., a quadrant photodiode.

FIGURE 5. *Intensity-based techniques for calibrating axial intensity decays*. (A), 'raisin cake', sub-diffraction fluorescent microspheres embedded in an index-matched (RI≈$n_1$) agarose gel. (B), oblique-fluorescent-layer sample, consisting of a fluorophore-coated coverslip and spacer (of height $d$, typically another cover glass). In a variant, (C), the surface of a long-$f$ lens of known curvature radius is coated with fluorescent beads and positioned on the interface and again the known axial fluorophore is used to probe the EW decay. In the 'infinitely' thick ($d \gg \lambda$) dye layer approach (D), a homogenous fluorophore solution is used to measure the *cumulative* fluorescence at a given penetration depth, $\delta(\theta)$. Upon multi-$\theta$ sweeps, depending on how far the EW reaches in the fluorescent solution, the intensity changes in a predictable manner. (E), in a dilute dye solution, the intensity fluctuations resulting from the diffusion of single molecules through the EW-excited volume allow measuring the penetration depth through TIR-FCS. (F), point emitters attached to the tip of an atomic force microscope (not shown) or attached to DNA probe the EW in a single point.

FIGURE 6. Refractive-index matched polymer height steps. (A), non-fluorescent polymer staircases, onto which sub-diffraction beads are drop cast to produce emitter layers at discrete distances, *left*, or covered with a dilute dye solution allow a tomographic reconstruction similar to that in Fig. 5D, *right*. (B), multi-layer sandwich of cell index-matched polymer spacer and capping layers (grey), separated by a thin fluorophore layer (green). (C), upon EW illumination at an angle $\theta$, samples with different fluorophore heights $\Delta$ (*top row*) result in different fluorescence intensities (*middle*), allowing the measurement of the axial intensity decay (*bottom*) by integrating the total fluorescence (*dashed circle*). Note the excitation spots at 3- and 9-o'clock positions on the BFP images. The refractive index $n_1$ of the polymer layers is obtained by SAF refractomertry {Brunstein, 2017 #59} from the radius at which the transition from under- to supercritical angle fluorescence occurs. The SAF/UAF ratio is directly proportional to *D*, {Oheim, 2018 #92}.

FIGURE 7. *Determining the effective penetration depth with a 'raisin-cake' test sample*. (A), experimental workflow. *Left*, acquisition of a *z*-stack of images in epifluorescence (EPI, $\theta = 0°$). *Middle*, EPI image taken at $z_0$ ($z := 0$, *red*), localizing the bottom layer of beads on the cover slip. *Right*, corresponding TIRF image at $z_0$ and different beam angles, $\theta$. (B), each bead was localized from its axial intensity profile by fitting a Gaussian (*black line*) with the average fluorescence, *F*(*z*), measured in a 3×3 px ROI (+). *Left,* example bead at *z* = 38 nm, with *F*(*z*) well described by a Gaussian. *Top* images correspond to planes identified by arrows. *Right*, example of a distorted profile discarded from analysis, *z* = 44 nm. (C), Intrinsic fluorescence, i.e., bead intensity measured from the peak of the Gaussian as in (B) for each bead, vs. its in-focus position *z*. (D), bead fluorescence $F^{(EPI)}(z; z_0)$ when focusing at the lowest bead layer at $z_0$, upon EPI illumination, as a function of previously measured bead position *z* and after normalization for its respective intrinsic fluorescence as in (C). The observed axial intensity decay (exponential fit, $D_z$ = 481±166 nm) is the result of the increasing defocus for surface-distant beads and the objective's finite depth of field. (D), same, for TIRF excitation (●) and fit of a double-exponential decay (*line*) with the measured fluorescence $F^{(TIRF)}(z; z_0)$ from all beads. Color codes for different $\theta$. The long-range component ($D_z$) was identical to that measured for out-of-focus beads upon EPI excitation, (D). The short-range component $\delta(\theta)$ was taken as the effective EW the penetration depth and was, respectively, 349±118, 139±20, 109±16 and 91±13 nm for $\theta$ = 65.0°, 67.5, 70.0°, and 72.5°. Note the surface-enhancement by a factor of ~4 of $F^{(TIRF)}(z_0)$ vs. $F^{(EPI)}(z_0)$. Depending on $\theta$, the fractional amplitude of the non-evanescent long-range excitation component varied between 13 and 17%, see main text. *Inset* shows double-exponential fits on a log scale and 95% confidence interval of the fit.

FIGURE S1. *Method for fixing beads on and above a glass coverslip*. Prior to the entire procedure a dash with a permanent marker pen was made on the upper face of the cover slip to facilitate focusing at the reflecting interface. Then, *left*, 1 μl of a 1:5 diluted solution of 0.1-μm diameter TetraSpeck[TM] was pipetted to the glass coverslip. Beads were left to dry so that they adhered to the glass. Subsequently, *middle*, 20 μl of agarose/sucrose solution again containing TetraSpeck[TM] beads at a 1:4 dilution was applied to the coverslip. The agarose was allowed to cool for polymerization. *Right*, the whole was topped with ~1.5 ml agarose/sucrose ($n_3 = 1.375$) solution previously kept at 35°C. After 10 min at 4°C the entire solution jellified. The result is a low-density carpet of beads attached to the coverslip, super-seeded with beads at different heights above the coverslip.

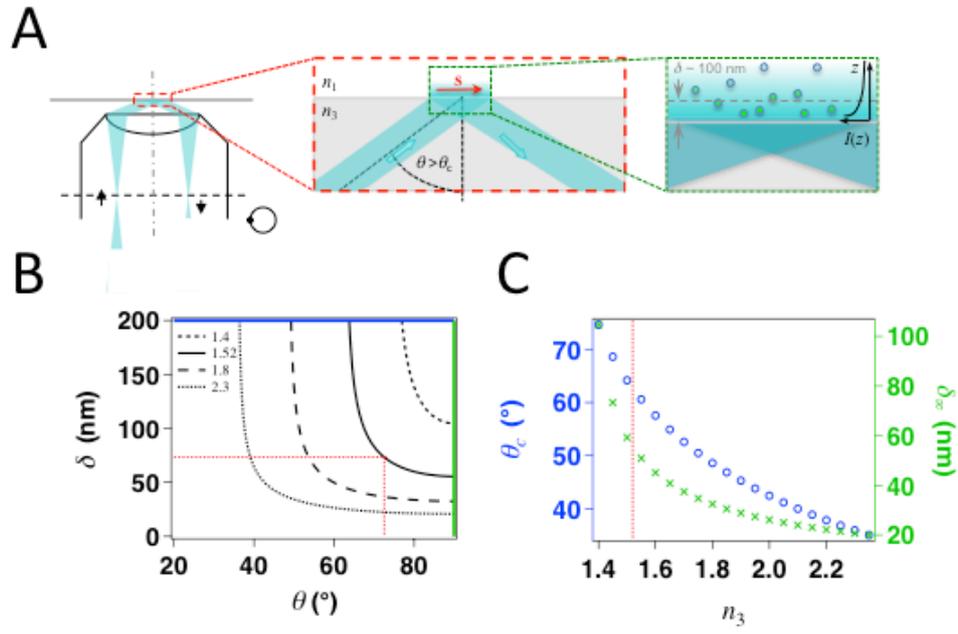

Oheim *et al.* Figure 1

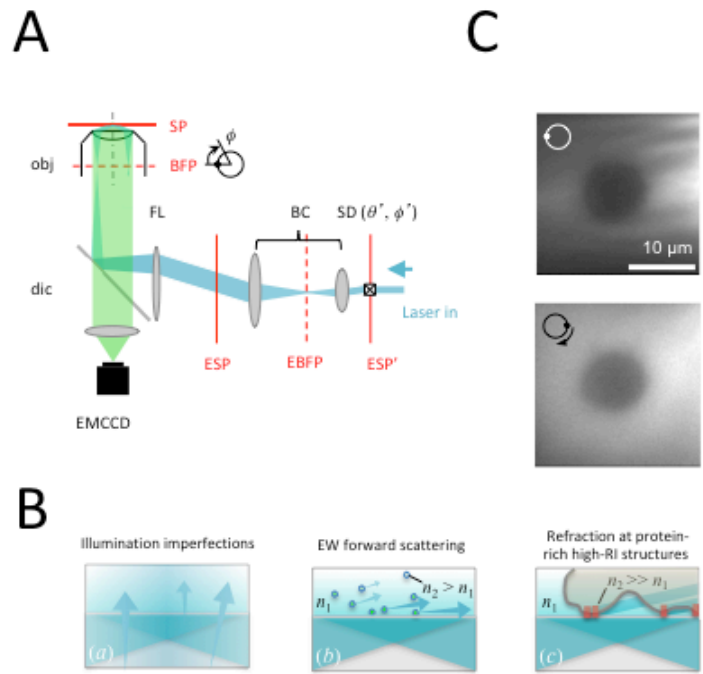

Oheim *et al.* Figure 2

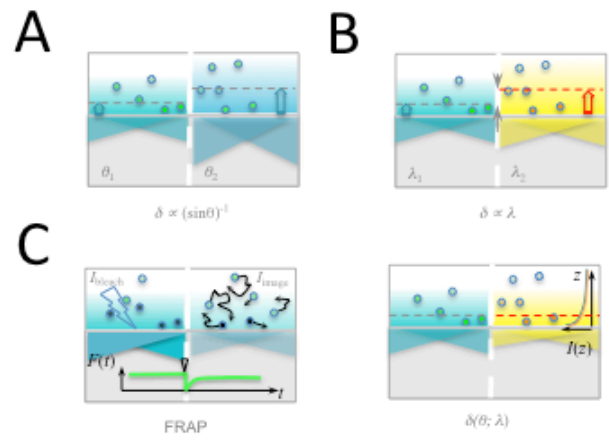

Oheim *et al.* Figure 3

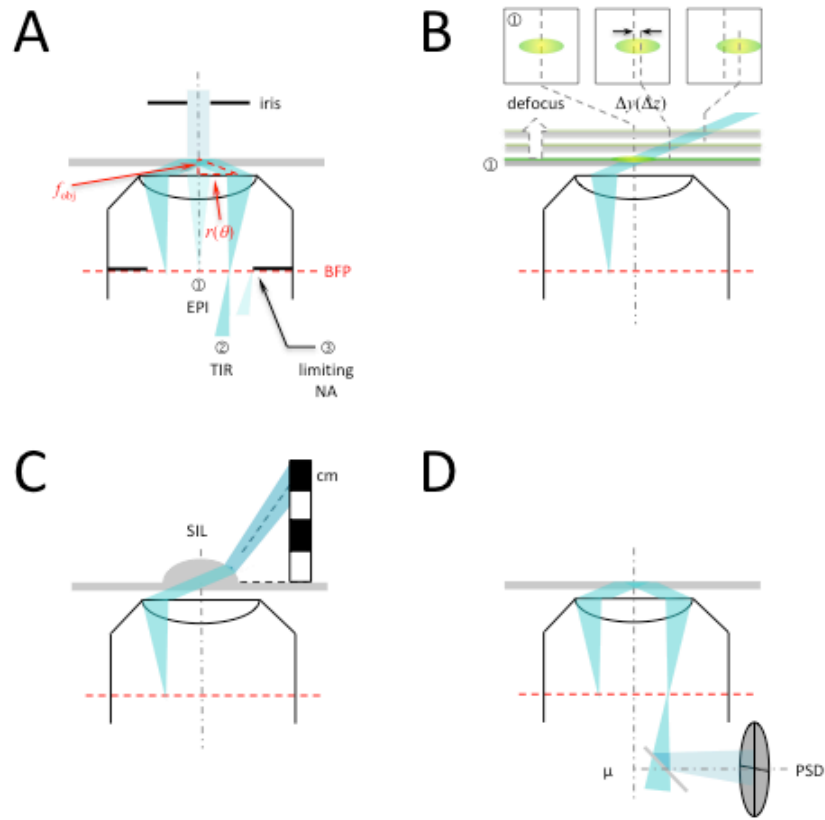

Oheim *et al.* Figure 4

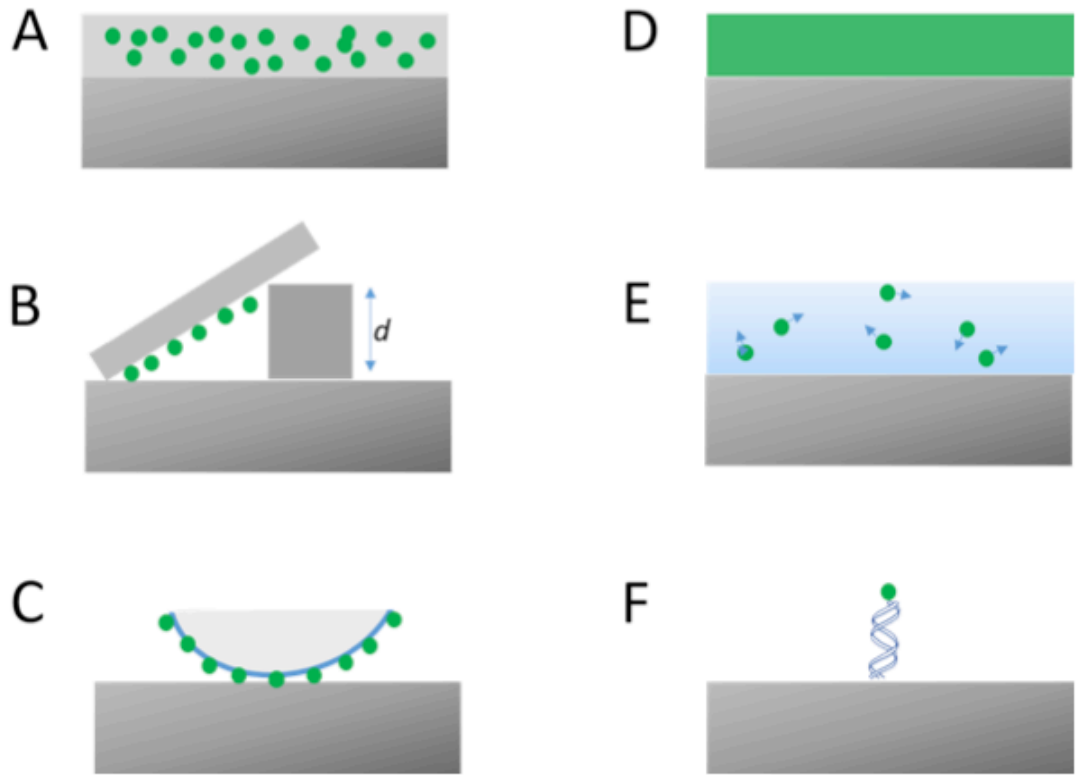

Oheim *et al.* Figure 5

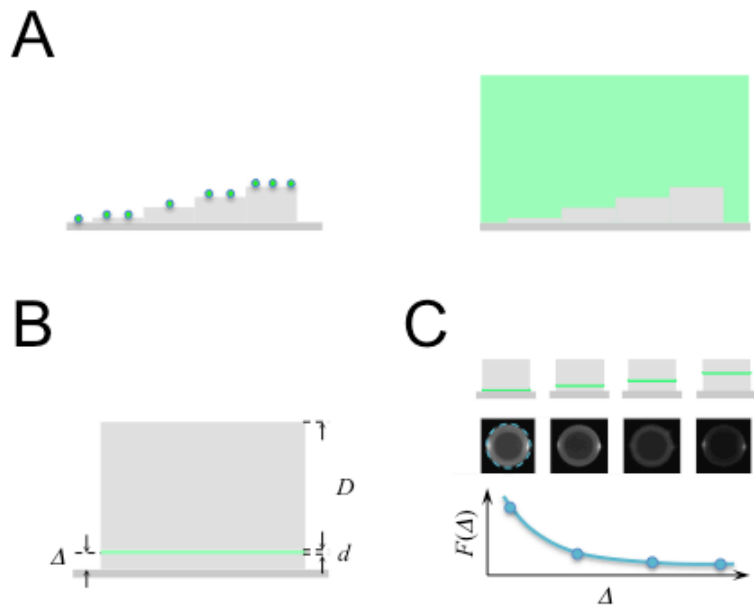

Oheim *et al.* Figure 6

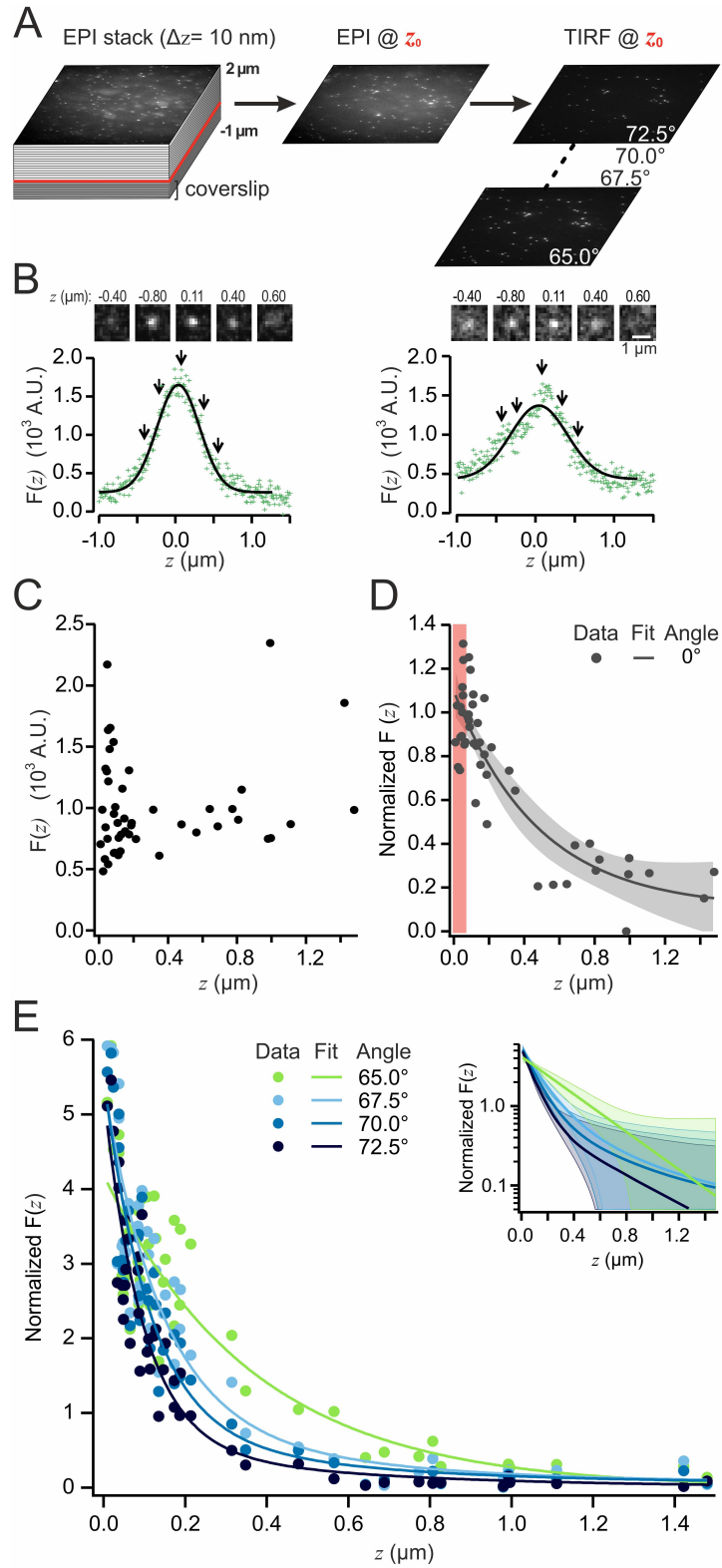

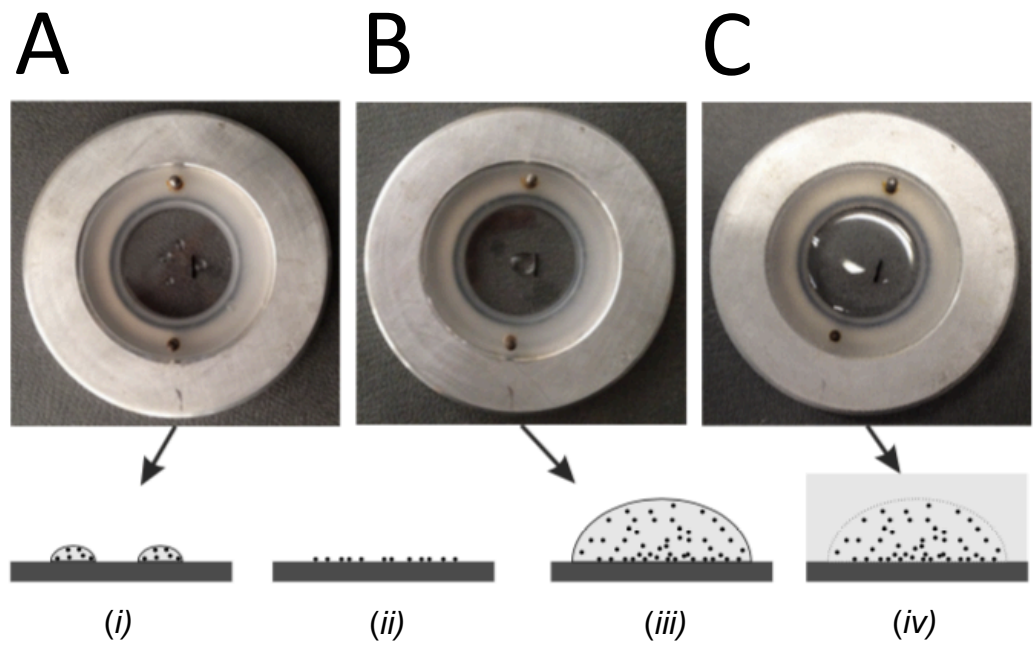

Oheim *et al.* Supplementary Figure S1